\documentclass[twocolumn,twocolappendix]{aastex631}

\usepackage{multirow}

\submitjournal{ApJ}

\shorttitle{Resolving High Redshift Galaxies with \textit{JWST}}
\shortauthors{Giménez-Arteaga et al.}

\begin{document}

\title{Spatially Resolved Properties of High Redshift Galaxies in the SMACS0723 \textit{JWST} ERO Field}

\correspondingauthor{Clara~Giménez-Arteaga}
\email{clara.arteaga@nbi.ku.dk}

\author[0000-0001-9419-9505]{Clara~Giménez-Arteaga}
\affiliation{Cosmic Dawn Center (DAWN), Denmark}
\affiliation{Niels Bohr Institute, University of Copenhagen, Jagtvej 128, DK-2200 Copenhagen N, Denmark}

\author[0000-0001-5851-6649]{Pascal~A.~Oesch}
\affiliation{Cosmic Dawn Center (DAWN), Denmark}
\affiliation{Niels Bohr Institute, University of Copenhagen, Jagtvej 128, DK-2200 Copenhagen N, Denmark}
\affiliation{Department of Astronomy, Université de Genève, Chemin Pegasi 51, CH-1290 Versoix, Switzerland}

\author[0000-0003-2680-005X]{Gabriel~B.~Brammer}
\affiliation{Cosmic Dawn Center (DAWN), Denmark}
\affiliation{Niels Bohr Institute, University of Copenhagen, Jagtvej 128, DK-2200 Copenhagen N, Denmark}

\author[0000-0001-6477-4011]{Francesco~Valentino}
\affiliation{Cosmic Dawn Center (DAWN), Denmark}
\affiliation{Niels Bohr Institute, University of Copenhagen, Jagtvej 128, DK-2200 Copenhagen N, Denmark}
\affiliation{European Southern Observatory, Karl-Schwarzschild-Str. 2, D-85748 Garching bei Munchen, Germany}

\author[0000-0002-3407-1785]{Charlotte~A.~Mason}
\affiliation{Cosmic Dawn Center (DAWN), Denmark}
\affiliation{Niels Bohr Institute, University of Copenhagen, Jagtvej 128, DK-2200 Copenhagen N, Denmark}

\author[0000-0001-8928-4465]{Andrea~Weibel}
\affiliation{Department of Astronomy, Université de Genève, Chemin Pegasi 51, CH-1290 Versoix, Switzerland}

\author[0000-0003-1641-6185]{Laia~Barrufet}
\affiliation{Department of Astronomy, Université de Genève, Chemin Pegasi 51, CH-1290 Versoix, Switzerland}

\author[0000-0001-7201-5066]{Seiji~Fujimoto}
\altaffiliation{Hubble Fellow}
\affiliation{Department of Astronomy, The University of Texas at Austin, Austin, TX 78712, USA}

\author[0000-0002-9389-7413]{Kasper~E.~Heintz}
\affiliation{Cosmic Dawn Center (DAWN), Denmark}
\affiliation{Niels Bohr Institute, University of Copenhagen, Jagtvej 128, DK-2200 Copenhagen N, Denmark}

\author[0000-0002-7524-374X]{Erica~J.~Nelson}
\affiliation{Department for Astrophysical and Planetary Science, University of Colorado, Boulder, CO 80309, USA}

\author[0000-0002-6338-7295]{Victoria~B.~Strait}
\affiliation{Cosmic Dawn Center (DAWN), Denmark}
\affiliation{Niels Bohr Institute, University of Copenhagen, Jagtvej 128, DK-2200 Copenhagen N, Denmark}

\author[0000-0002-1714-1905]{Katherine~A.~Suess}
\affiliation{Department of Astronomy and Astrophysics, University of California, Santa Cruz, 1156 High Street, Santa Cruz, CA 95064 USA}
\affiliation{Kavli Institute for Particle Astrophysics and Cosmology and Department of Physics, Stanford University, Stanford, CA 94305, USA}

\author[0000-0003-1903-9813]{Justus~Gibson}
\affiliation{Department for Astrophysical and Planetary Science, University of Colorado, Boulder, CO 80309, USA}

\begin{abstract}

We present the first spatially resolved measurements of galaxy properties in the \textit{JWST} ERO SMACS0723 field. We perform a comprehensive analysis of five $5<z<9$ galaxies with spectroscopic redshifts from NIRSpec observations and photometric coverage with NIRCam. We perform spatially resolved spectral energy distribution fitting with \textsc{Bagpipes}, using NIRCam imaging in 6 bands spanning the wavelength range 0.8$-$5$\mu$m. We produce maps of the inferred physical properties by using a novel approach in the study of high redshift galaxies. This method allows us to study the internal structure and assembly of the first generations of galaxies. We find clear gradients both in the empirical colour maps, as well as in most of the estimated physical parameters. We find regions of considerably different specific star formation rates across each galaxy, which points to very bursty star-formation happening on small scales, not galaxy-wide. The integrated light is dominated by these bursty regions, which exhibit strong line emission detected by NIRSpec and also inferred from the broad-band NIRCam images, with the equivalent width of [\ion{O}{3}]+H$\beta$ reaching up to $\sim3000-4000$~Å rest-frame in these regions. Studying these galaxies in an integrated approach yields extremely young inferred ages of the stellar population ($<$10~Myr), which outshine older stellar populations that are only distinguishable in the spatially resolved maps. This leads to inferring $\sim0.5-1$ dex lower stellar masses by using single-aperture photometry, when compared to resolved analyses. Such systematics would have strong implications in the shape and evolution of the stellar mass function at these early times, particularly while samples are limited to small numbers of the brightest candidates. Furthermore, the evolved stellar populations revealed in this study imply an extended process of early galaxy formation that could otherwise be hidden behind the light of the most recently formed stars.

\end{abstract}


\section{Introduction} \label{sec:intro}

By characterizing the physical properties of galaxies in the redshift range $5<z<10$, we can study the epoch of reionization, when the Universe experienced its last phase transition \citep[see e.g.,][for a review]{Treu13,Mason18,Robertson22}. With well-sampled photometry of high redshift galaxies, we can robustly model their spectral energy distributions (SEDs) and infer the properties of their stellar populations. Up until now, the rest-frame optical emission from galaxies was unavailable at $z>7$, having been redshifted to the part of the near-infrared spectrum where our facilities lacked sensitivity and spatial resolution. While the rest-frame UV emission we have had access to is a good tracer of unattenuated star formation, it is a poor tracer of stars older and less massive than O and B-type, that make up the bulk of total stellar mass for populations older than a few Myr.

The latest addition to the space fleet of telescopes, the \textit{James Webb Space Telescope} (\textit{JWST}), has unprecedented sensitivity and spatial resolution in the near infrared. This has opened up a new window into the rest-frame optical emission of high redshift galaxies, allowing us to understand their stellar populations for the first time. The Near-Infrared Camera (NIRCam; \citealt{2005SPIE.5904....1R}) on board \textit{JWST} allows us to reach this spectral range with a unique depth and resolution. The Near Infrared Spectrograph (NIRSpec; \citealt{NIRSpec}) provides high resolution spectroscopy in the near-infrared, which is key to robustly determine the redshift. This improves the modelling of the SEDs by constraining a free parameter, thus breaking the degeneracies that the redshift has with age and dust \citep[see e.g.,][for a review]{Conroy13}.

Lower-redshift studies have been able to resolve galaxies and their components \citep[up to $z\sim2$ in e.g.,][]{Zibetti09,Nelson19,Morselli19,Suess19,Abdurrouf21,Arteaga22}. 
At higher redshifts, resolved analyses have typically only been possible in lensed systems \citep[e.g.,][]{Zitrin11,Vanzella17}, or in particularly luminous galaxies that break up into multiple components \citep[e.g.,][]{Matthee20,Bowler22}.
Nevertheless, integrated photometry has revealed the population demographics: stellar mass functions, number counts \citep[e.g.,][]{Song16,Stefanon15,Stefanon21}. 

With \textit{JWST} we can extend for the first time resolved studies beyond redshift $\sim2$, introducing the possibility to study in unique detail the first generations of galaxies \citep[e.g.,][]{Chen22,Hsiao22}. These resolved studies will allow us to place unique, new constraints on the formation and evolution of the first galaxies: their mass assembly histories, modes of growth, chemical enrichment, and earliest quenching mechanisms. In order to build a complete picture of galaxy assembly, a resolved view of its components is required, to fully understand the interplay between the stellar population, dust and gas in $z>6$ galaxies.

The impact of having resolved observations has so far only been studied at low redshifts ($z\lesssim3$). Various works have compared the inferred physical properties obtained with resolved and unresolved observations \citep[see e.g.,][]{Wuyts12,Sorba15,Sorba18,ValeAsari20,Fetherolf20}, with diverse conclusions. A resolved approach can have multiple advantages, such as decreasing degeneracies in the stellar population synthesis models and producing more realistic star formation histories \citep[][]{PerezGonzalez22}. In highly star forming galaxies, the outshining of old stellar populations by young ones is of particular importance, which a resolved analysis could untangle. \cite{Sorba18} find that the total stellar mass can be underestimated by a factor of $\sim5$ without taking this outshining effect into account with spatially-resolved SED modelling. 

In this paper we present the observations and first results of a multi-wavelength analysis of five high redshift galaxies ($5 < z < 9$) observed with \textit{JWST}, to study the spatially-resolved properties of their stellar populations. These galaxies are the highest spectroscopically confirmed targets in the \textit{JWST} ERO SMACS0723 field. Using NIRCam imaging in 6 bands spanning the wavelength range $0.8-5 \mu$m, we perform spatially-resolved SED fitting with \textsc{Bagpipes} \citep{bagpipes}. There have been multiple works on these targets, albeit always from an integrated perspective \citep[e.g.,][]{Carnall22,Tacchella22,Schaerer22,Trump22,Rhoads22,Curti22,Brinchmann22,Arellano22,Heintz22,Fujimoto22}. These papers derive different stellar masses, some of them with extremely early star formation histories. Here we present the first spatially resolved analysis on these high redshift galaxies, which we propose as a more robust approach to accurately calculate their stellar masses. 

This paper is structured as follows. In Section \ref{sec:data}, we introduce the \textit{JWST} observations and data reduction procedure. Section \ref{sec:method} describes the methodology we use for the modelling of the SEDs with \textsc{Bagpipes}. In Section \ref{sec:results}, we present the main results and inferred properties of our sample, both in integrated and resolved approaches, as well as discussing the implications of our analyses. Finally, in Section \ref{sec:conclusions} we present the summary and conclusions of our work. Throughout this paper, we assume a simplified $\Lambda$CDM cosmology with $H_0 = 70$ km s$^{-1}$ Mpc$^{-1}$, $\Omega_m$= 0.3 and $\Omega_{\Lambda} = 0.7$. No lensing correction is applied throughout this work. Hence, intrinsic stellar masses and star formation rates can be obtained by dividing by the magnification factor ($\mu$, see Table~\ref{tab:sample}).

\section{Data and Observations} \label{sec:data}

\begin{figure*}[t]
\centering
\includegraphics[width=\textwidth]{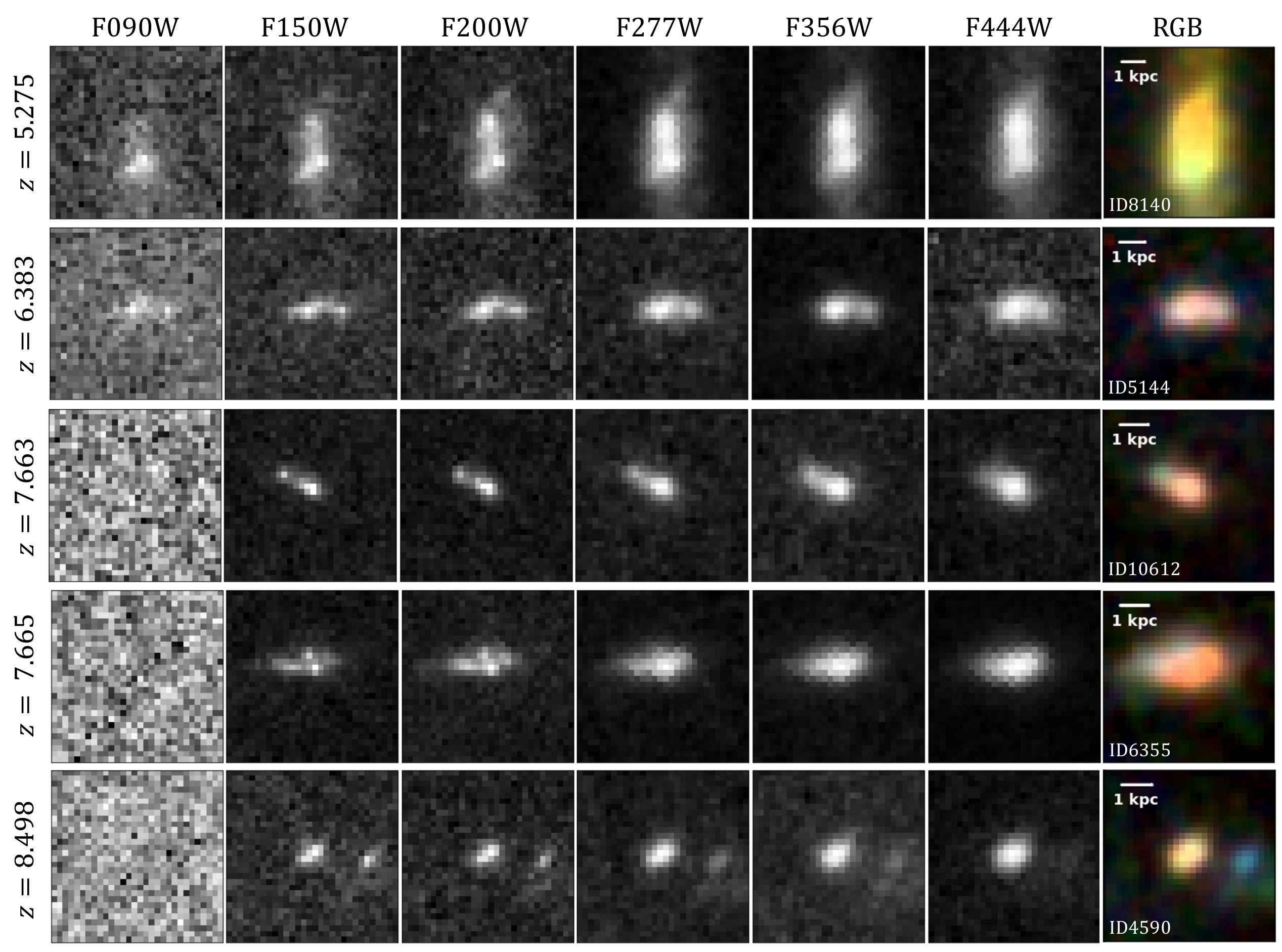}
\caption{Cutout images of the five SMACS0723 galaxies in all available NIRCam bands. The cutouts are 1\farcs2 across and centered in the coordinates provided in Table \ref{tab:sample}. The RGB images are built combining the F150W (B), F277W (G) and F444W (R) PSF-matched images (to the F444W band). The physical scale is calculated in the lens plane. \label{fig:cutouts}}
\end{figure*}

We use the public data of the galaxy cluster SMACS J0723.3-7327 (SMACS0723), observed by \textit{JWST} as part of the Early Release Observations (ERO; Programme ID 2736, \citealt{2022ApJ...936L..14P}). We use the Near-Infrared Camera (NIRCam; \citealt{2005SPIE.5904....1R}) photometric data from the catalog reduced by Brammer et al. (in prep). The data has been reduced using the public software package \texttt{grizli} \citep{Brammer19,grizli,grizli22}.  The photometry is corrected for Milky Way extinction assuming $E(B-V) = 0.1909$ \citep{MW11} and the \cite{Fitzpatrick07} extinction curve. The images are PSF-matched to the F444W band on a common 0\farcs04/pixel scale. We adopt the PSF models for use with the \texttt{grizli} mosaics\footnote{\url{https://github.com/gbrammer/grizli-psf-library/tree/main/smacs0723}}, which are based on the \texttt{WebbPSF} models. We compute matching kernels for each of the PSFs to the F444W PSF using a Richardson-Lucy deconvolution algorithm \citep{Richardson72,Lucy74}, and then convolve the images with the resulting kernels to match the PSF resolution in F444W.

In this work we focus on the five galaxies at $5<z<9$ that have spectroscopic redshifts confirmed from NIRSpec observations, presented in \cite{Carnall22}. These are detected in the 6 deep NIRCam imaging filters F090W, F150W, F200W, F277W, F356W, and F444W. The targets also have shallower imaging data obtained with NIRISS with the F115W and F200W filters, which we exclude in the analysis presented in this work due to their lower resolution. The sample spans a redshift range from 5.275 for the closest galaxy, up to 8.498 for the highest redshift object \citep{Carnall22}. The sources have magnification factors between 1.6 up to 10.1 in the \textsc{Glafic} lens models from \cite{Oguri10}. They are not significantly distorted by the gravitational lensing so that it affects the spatially resolved SED fitting. Table \ref{tab:sample} provides the basic information for the targets. We do not apply any lensing correction to our results, although we report on Table~\ref{tab:sample} the lensing factors for these sources presented in \cite{Carnall22}. Figure \ref{fig:cutouts} displays the cutout images of the five galaxies in all available observed bands, as well as the colour images built combining the F150W, F277W, and F444W filters.

\begin{table}
\centering
\begin{tabular}{l c c c c}
\hline
Redshift & ID & RA & DEC & Lensing $\mu$ \\ 
\hline
 5.275 & 8140 & 110.78804 & $-$73.46179 & 1.7 \\
 6.383 & 5144 & 110.83972 & $-$73.44536 & 2.9 \\
 7.663 & 10612 & 110.83395 & $-73.43454$ & 1.6 \\
 7.665 & 6355 & 110.84452 & $-$73.43508 & 2.7 \\
 8.498 & 4590 & 110.85933 & $-$73.44916 & 10.1 \\
\hline
\end{tabular}
\caption{Redshift, coordinates and magnification factors for the 5 high-redshift galaxies studied in this work. The information is from \cite{Carnall22}, where the lensing factors ($\mu$) are taken from \cite{Oguri10}.}
\label{tab:sample}
\end{table}

%%%%
\section{Methodology} \label{sec:method}

\subsection{SED fitting with \textsc{Bagpipes}}\label{sec:sed_bagpipes}
To model the spectral energy distribution of the individual pixels and derive the physical properties, we use the SED fitting code \textsc{Bagpipes} \citep[][]{bagpipes}. We set the NIRSpec spectroscopic redshifts indicated on Table~\ref{tab:sample}, in order to break the degeneracy with age and dust that a photometric or uncertain redshift would introduce. We use the SPS models by \cite{BC03} and include the nebular emission with \textsc{Cloudy} \citep[][]{cloudy}, extending the \textsc{Bagpipes} default grid (that normally reaches up to $\log_{10}(U)=-2$) so that the ionization parameter, $U$, varies from $-3 < \log_{10}(U) < -1$, since $z>6$ galaxies display higher ionization parameters than low-redshift galaxies \citep[e.g.,][]{Sugahara22,Curti22}. We assume a \cite{Kroupa01} initial mass function (IMF), and a \cite{Calzetti00} attenuation curve, in order to reduce the number of free parameters in our fits. We choose a constant star formation history model, following \cite{Carnall22}, which seems adequate to fit our galaxies (we obtain reduced $\chi^2$ values within the range $0.1-7.5$, shown in further detail in the following sections). The formation of very young, low-mass, and low-metallicity galaxies is likely bursty, and a constant SFH accurately resembles this on short timescales. We let the maximum age grid to vary from 1~Myr to 1~Gyr, to allow for the presence of more evolved stellar populations, and further limited by the age at the given redshift. We set the visual extinction to vary from $A_V=0$ to $A_V=2$, and the metallicity from 0 to $Z_{\odot}$, with uniform priors. Even though the metallicity has been calculated with integrated NIRSpec spectra for these targets \citep[see e.g.,][]{Schaerer22,Curti22,Brinchmann22}, they obtain varying results within our allowed range. Moreover, we want to allow for spatial variation across the galaxy, thus we do not set $Z$ as a fixed parameter in the fit. Finally, we set the lifetime of birth clouds to 10~Myr.

\subsection{Pixel-based modelling}

In this work we perform SED fitting on a pixel-by-pixel basis. With the setup described in the previous subsection, we fit the spectral energy distribution and infer the physical parameters of each individual pixel. This allows us to recover the 2D distribution of properties such as the stellar mass and star formation rate (SFR). The pixels in our maps correspond to physical sizes between 180 and 240 parsecs. In order to fit the SED of individual pixels, we impose a signal-to-noise ratio (S/N) threshold of 2 on both the F150W and F200W bands, which are the noisiest. We find that this threshold is enough to produce trustworthy fits, obtaining good reduced $\chi^2$ values in the fits of individual pixels, as we show in more detail in the following section. To produce the maps of the physical properties and study their spatial distribution, we display the 50th percentile of the inferred parameter, calculated with the posterior distribution that \textsc{Bagpipes} provides. We can also present the uncertainties for each pixel extracted from the 16th and 84th percentiles of the posterior distribution (see \citealt{bagpipes} for details).

%%%%
\section{Results and Discussion} \label{sec:results}

In this section we present and discuss the results of our study. We provide both an integrated and a spatially resolved analysis of the physical properties that we infer for the five targets that comprise this work.

\subsection{Spatially Resolved Physical Properties} \label{sec:results_SR}

\begin{figure*}[t]
\centering
\includegraphics[width=\textwidth]{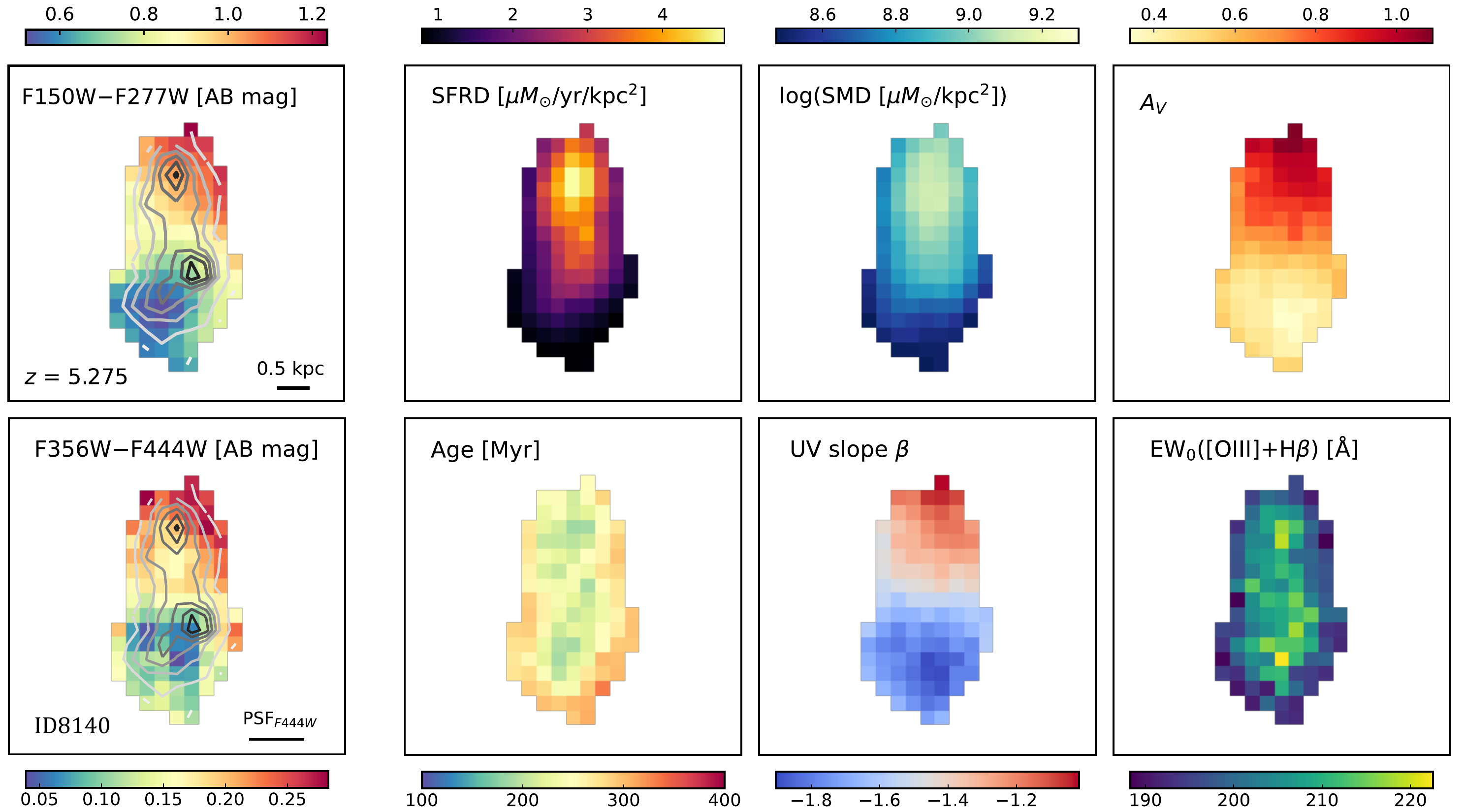}
\caption{Maps of the galaxy ID8140 at $z=5.275$. \textbf{Left:} Maps of the empirical colours in AB mag, inferred as the difference of the bands F150 and F277W (top) and the images F356W and F444W (bottom). The contours correspond to the non PSF-matched F200W image. The physical scale is calculated in the lens plane. The FWHM of the F444W PSF is indicated. \textbf{Right:} Maps of the physical properties inferred with \textsc{Bagpipes}. The top row, from left to right, shows the resulting maps for the star formation rate surface density (SFRD), the logarithm of the stellar mass surface density (SMD), and the visual extinction $A_V$. The bottom row shows the maps for the mass-weighted stellar age, the UV slope ($\beta$) and the equivalent width of the inferred [\ion{O}{3}]+H$\beta$ emission lines. No lensing correction has been applied.}
\label{results_z5.275}
\end{figure*}

For all galaxies studied, we have produced maps of the physical parameters inferred with \textsc{Bagpipes}. These include the star-formation rate surface density (SFRD), the stellar mass surface density (SMD), the visual extinction $A_V$, the mass weighted age, the UV slope $\beta$, and the equivalent width (EW) of the [\ion{O}{3}] and H$\beta$ emission lines. The last two are measured from the \textsc{Bagpipes} posterior SEDs. We also display maps of the empirical colours F150W--F277W, which is a proxy for the UV slope, and F356W--F444W, which generally traces the strength of the inferred [\ion{O}{3}]+H$\beta$ emission (except for the lowest and highest redshift galaxies, where no available couple of bands capture the lines and continuum accordingly).

\begin{figure*}[!th]
\centering
\includegraphics[width=\textwidth]{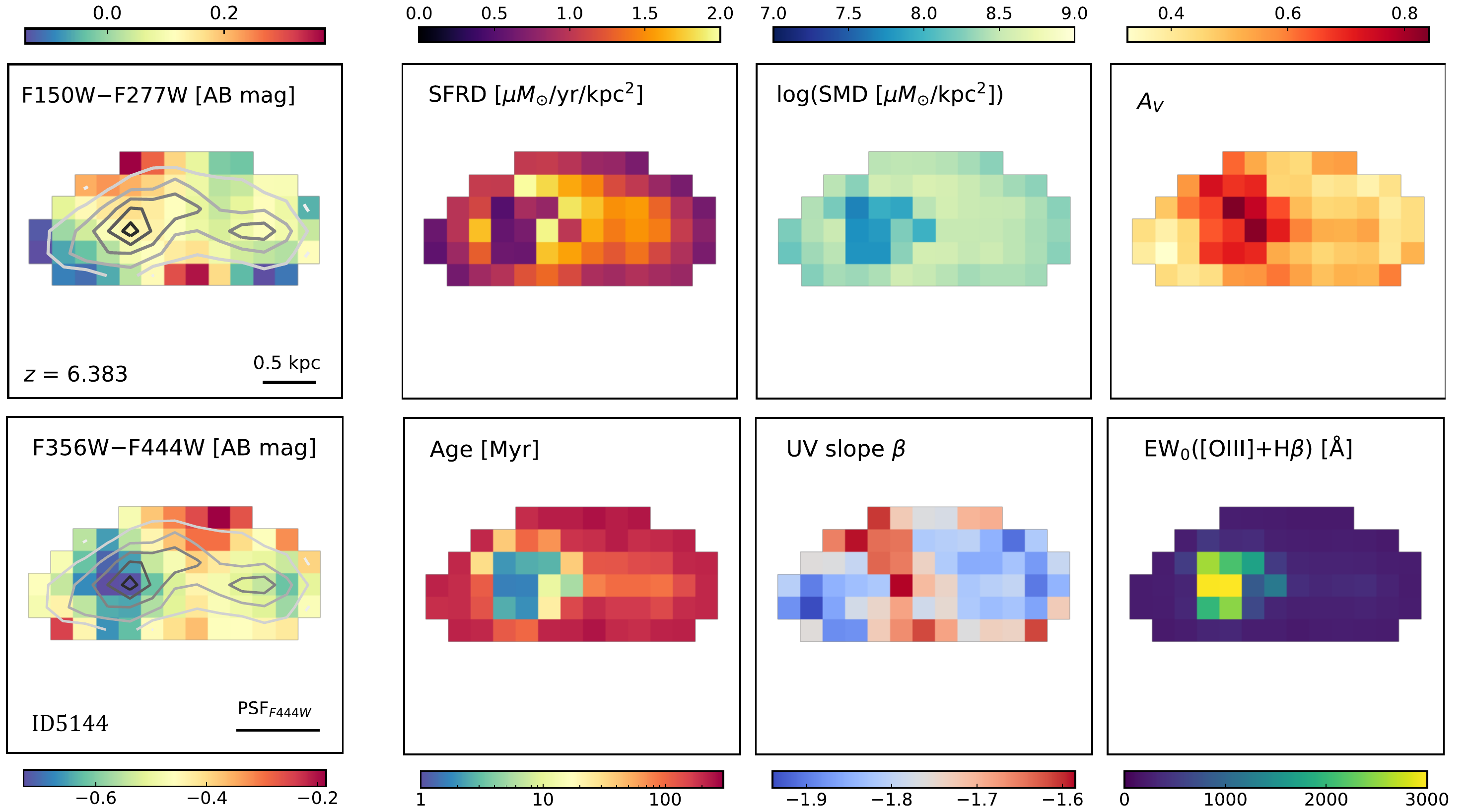}
\caption{Maps of the empirical colours and the physical properties inferred with \textsc{Bagpipes} on the galaxy ID5144 at $z=6.383$. See Figure \ref{results_z5.275} for more details.} 
\label{results_z6.383}
\end{figure*}

\begin{figure*}[!th]
\centering
\includegraphics[width=\textwidth]{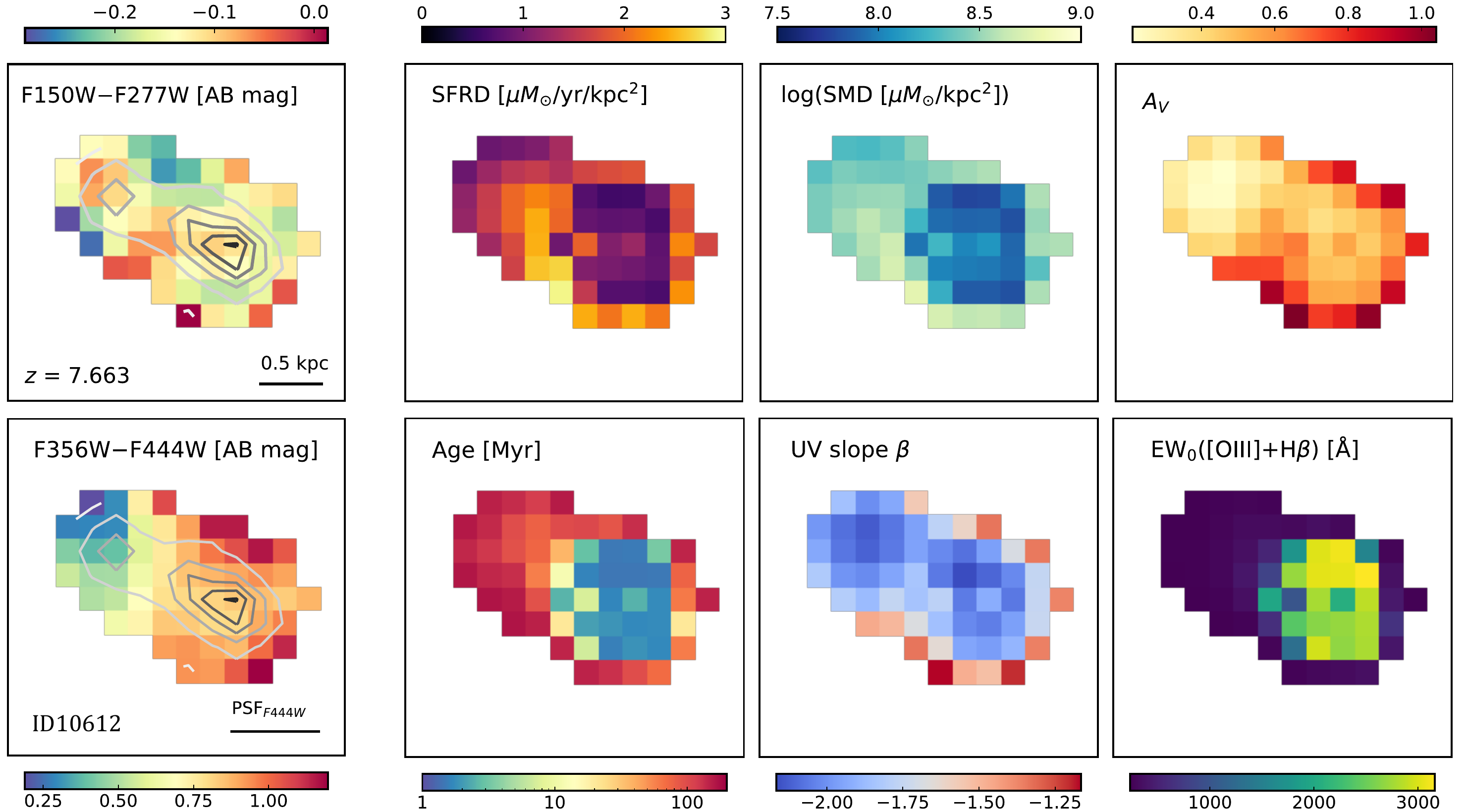}
\caption{Maps of the empirical colours and the physical properties inferred with \textsc{Bagpipes} on the galaxy ID10612 at $z=7.663$. See Figure \ref{results_z5.275} for more details.} 
\label{results_z7.663}
\end{figure*}

\begin{figure*}[!th]
\centering
\includegraphics[width=\textwidth]{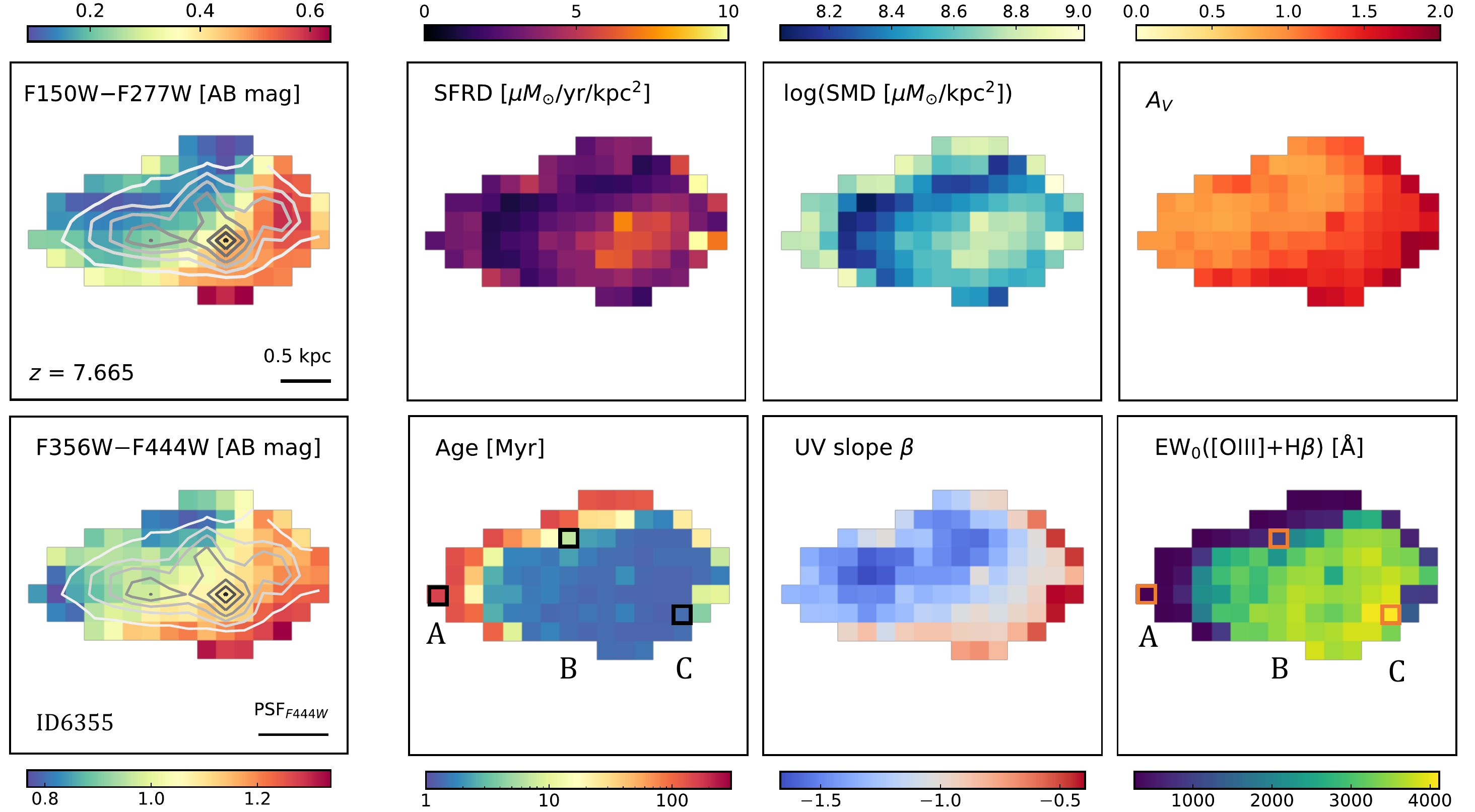}
\caption{Maps of the empirical colours and the physical properties inferred with \textsc{Bagpipes} on the galaxy ID6355 at $z=7.665$. See Figure \ref{results_z5.275} for more details. The three boxes A, B, C indicate the pixels that are analysed in more detail in the text, as well as in Figure~\ref{results_individual_seds}, where the best fit SEDs are shown for each individual pixel.} 
\label{results_z7.665}
\end{figure*}

\begin{figure*}[!th]
\centering
\includegraphics[width=\textwidth]{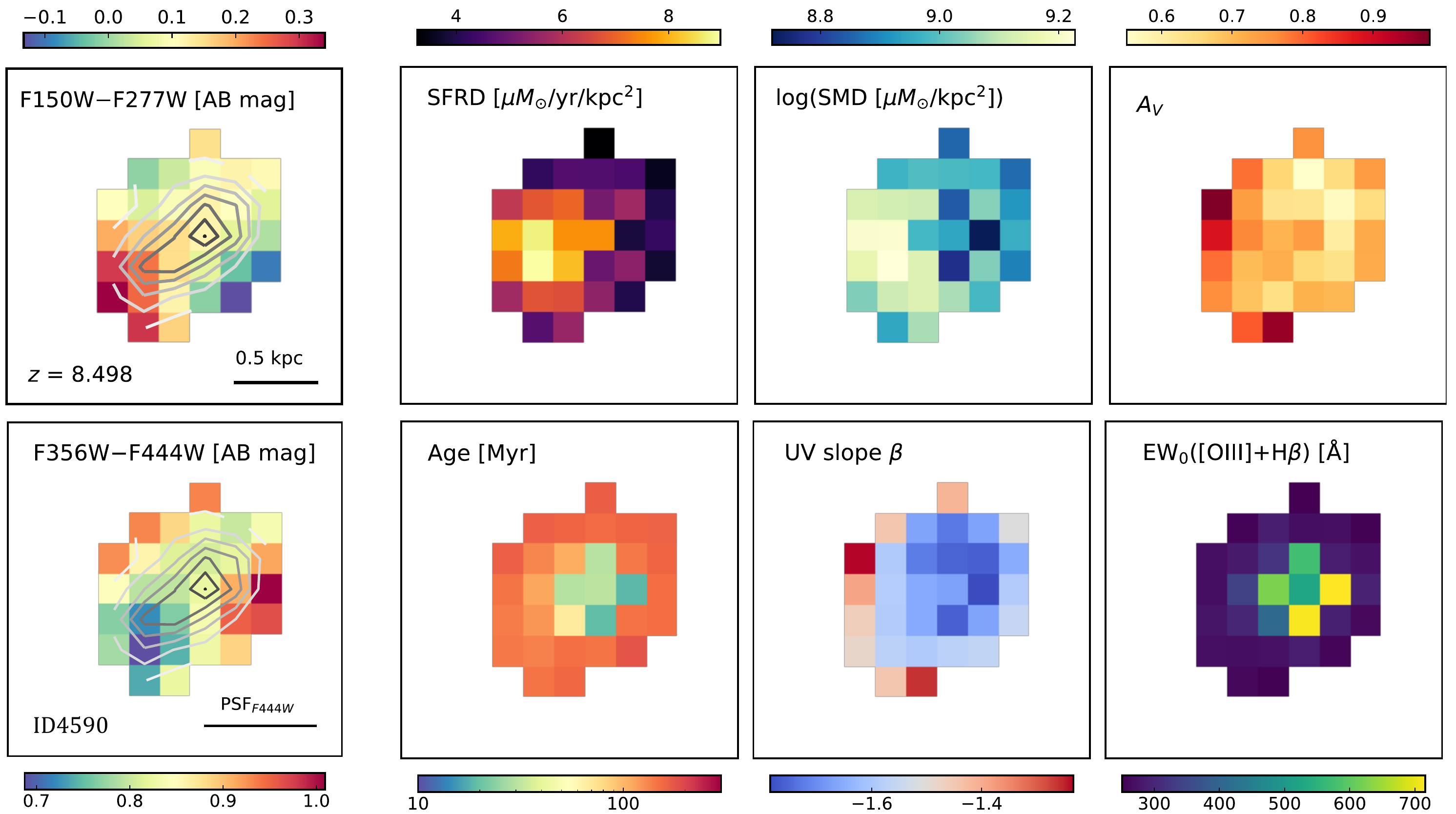}
\caption{Maps of the empirical colours and the physical properties inferred with \textsc{Bagpipes} on the galaxy ID4590 at $z=8.498$. See Figure \ref{results_z5.275} for more details.} 
\label{results_z8.498}
\end{figure*}

Figures~\ref{results_z5.275}--\ref{results_z8.498} display the resulting maps for the five galaxies presented in this work, from the lowest redshift $z=5.275$, to the highest $z=8.498$. Firstly, we see that all galaxies are resolved and display strong empirical colour gradients, both in the blue bands as well as the red bands. These gradients appear on larger scales than the FWHM of the F444W PSF (0\farcs145, equivalent to $\sim3.6$ pixels), confirming that we can resolve trends and structures in our sample. In general, we find that even at these early times, most of the galaxies display multiple star forming clumps, traced by the F200W contours, as is found also by other recent works \citep[e.g.,][]{Claeyssens22,Treu22, Chen22}. These are regions of very high inferred equivalent widths of the [\ion{O}{3}]+H$\beta$ emission (in the range $\sim300-4000$~Å rest-frame), embedded within larger structures that are not undergoing a burst of star formation. In these regions with extreme line EWs, the inferred ages are extremely young ($<$10~Myr), corresponding to a bursty clump of young stars that is resolved in targets ID10612 (Figure~\ref{results_z7.663}) and ID6355 (Figure~\ref{results_z7.665}), and marginally unresolved in ID5144 (Figure~\ref{results_z6.383}) and ID4590 (Figure~\ref{results_z8.498}), given the scale of the F444W PSF FWHM. Around these high-EW bursty clumps, we also find underlying older stellar populations ($\sim$100~Myr), which would be missed in an integrated analysis, as we will discuss in more detail in the next subsection.

Figure \ref{results_z5.275} displays the maps for galaxy ID8140, at redshift $z=5.275$. We obtain smooth maps for all physical properties and colours. The SFRD and SMD appear entirely co-spatial, and the UV slope traces perfectly the dust obscuration map. This galaxy has $\sim2$ times older stellar populations compared with the average of the rest of the sample, in line with being the galaxy at lowest redshift. It shows strong colour, $A_V$ and UV slope gradients, with two distinct clumps, one red and one blue. This could indicate that this galaxy is undergoing a merger, even though the EWs (the lowest of all targets) and ages show little variation across the object.

\begin{figure}[!th]
\centering
\includegraphics[width=\columnwidth]{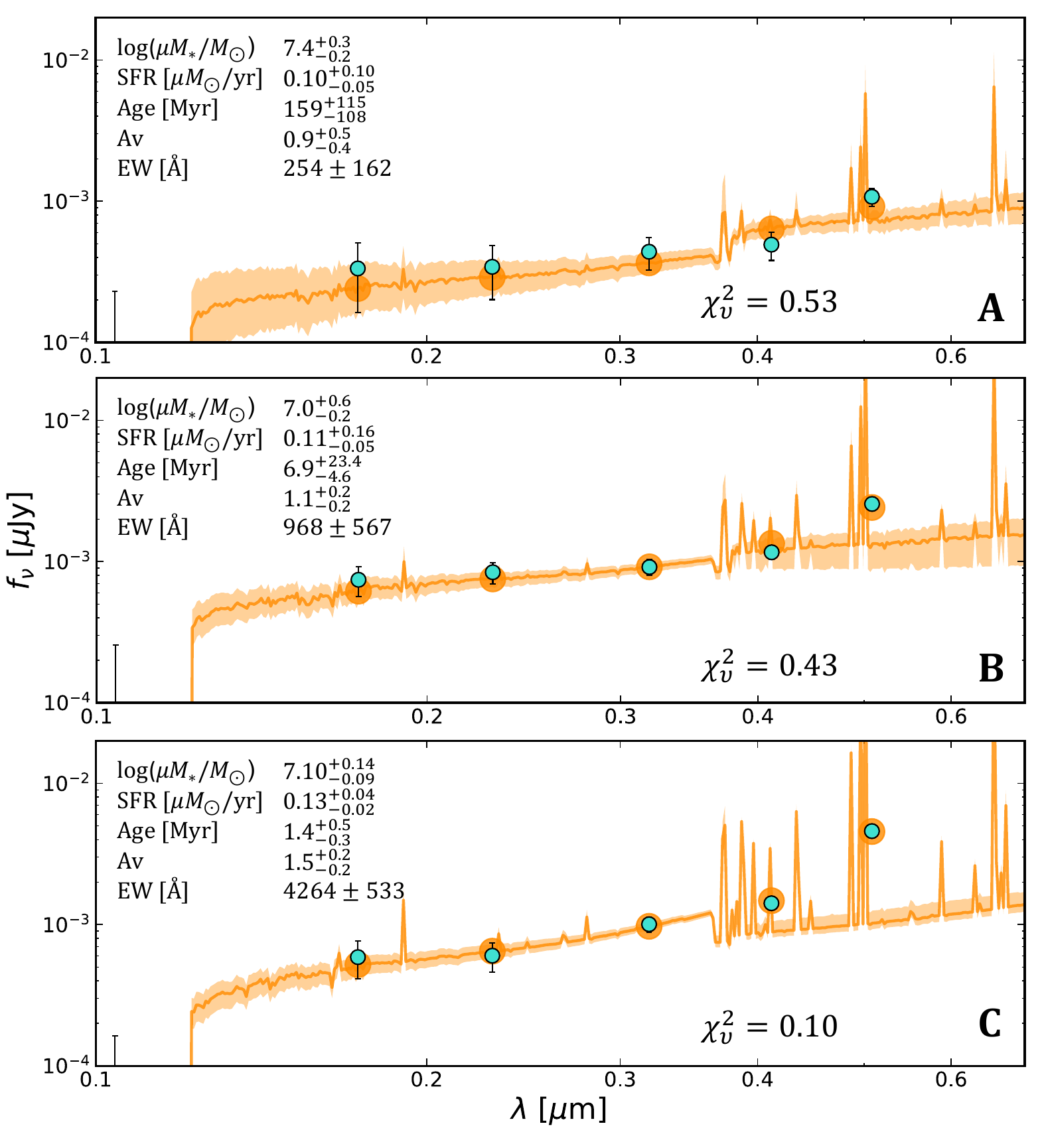}
\caption{Best fit SEDs for the three pixels A, B and C indicated in Figure~\ref{results_z7.665}, from the galaxy ID6355 at $z=7.665$. The turquoise points and errorbars correspond to the NIRCam photometry, the orange points to the best fit model, and the orange curve and shaded region is the best fit SED inferred with \textsc{Bagpipes}, and corresponding 16th and 84th percentile uncertainty interval. The inferred physical parameters are indicated for each pixel, as well as the reduced $\chi^2$ of the fit.}
\label{results_individual_seds}
\end{figure}

Figure~\ref{results_z7.665} shows the resulting maps for the galaxy ID6355 at $z=7.665$. It is the galaxy with most extreme EWs (reaching $\sim4000$~Å). We see that the region with very high EW is very extended for this galaxy, leaving barely a shell where we find underlying older stellar populations, which are otherwise outshined (or not present) by the younger stars in these strong line emission regions. The clear gradients in the empirical colours give us confidence that this shell is real and not an artifact of the age-dust degeneracy in the SED fitting process, which we also discuss in further detail in \S\ref{section:caveats}. On top of this, the shell is larger than the PSF scale. In Figure~\ref{results_individual_seds} we present and analyse the fits for three individual pixels within this source, so that we can study further whether the ``shell" of older stars in this particular galaxy is real or an artifact. We select the pixels A, B, and C that are indicated in the age and EW maps in Figure~\ref{results_z7.665}, since they appear to be very distinct regions within this galaxy. Albeit being towards the edge of the galaxy, all three pixels fulfill our S/N threshold, so that we can produce robust fits (with reduced $\chi^2$ values within $0.10-0.53$). These pixels are also far enough from each other so that the PSF is not blending the information they encode, and we can thus resolve their different stellar populations. We can clearly see that the SEDs look different, reflecting the gradients that we already see in Figure~\ref{results_z7.665}, both on the maps of the inferred physical parameters, as well as the empirical colour maps. The greatest difference is observed in the strength of the inferred [\ion{O}{3}]+H$\beta$ emission lines, since the SED for pixel C has extreme EW, reaching $4264\pm533$~Å rest-frame. This yields a considerable difference in the inferred ages, with pixel A having a mass weighted age of $159^{+115}_{-108}$~Myr, and pixels B and C displaying very young stellar ages under 10~Myr.

Figure~\ref{results_z8.498} shows the results for the galaxy ID4590, which is the highest redshift $z=8.498$ in this work. It is a very compact source, with only 31 pixels where S/N$>$2 for both F150W and F200W. We see a star forming clump, which also corresponds to the highest inferred stellar mass, and towards the dustiest zone in the $A_V$ map. We see a marginally unresolved centrally located clump of young stellar population, which is not entirely co-spatial with the star-forming burst, but traces perfectly the higher equivalent width of the inferred [\ion{O}{3}] and H$\beta$ emission lines. On top of this, the empirical colour maps display a clear gradient, which follows the ones observed for the SFRD, SMD, $A_V$ and $\beta$ maps. The rest of targets exhibit similar trends for all physical properties, with the main characteristic being this region with extremely high EWs and therefore very young stellar populations.

\subsection{Integrated Analysis}\label{sec:integrated}

Besides providing an invaluable insight into the internal structure of galaxies, we want to test whether spatially resolved observations yield other consequences, such as inferring different physical properties, compared to only integrated measurements. 

\begin{figure}[t]
\centering
\includegraphics[width=\columnwidth]{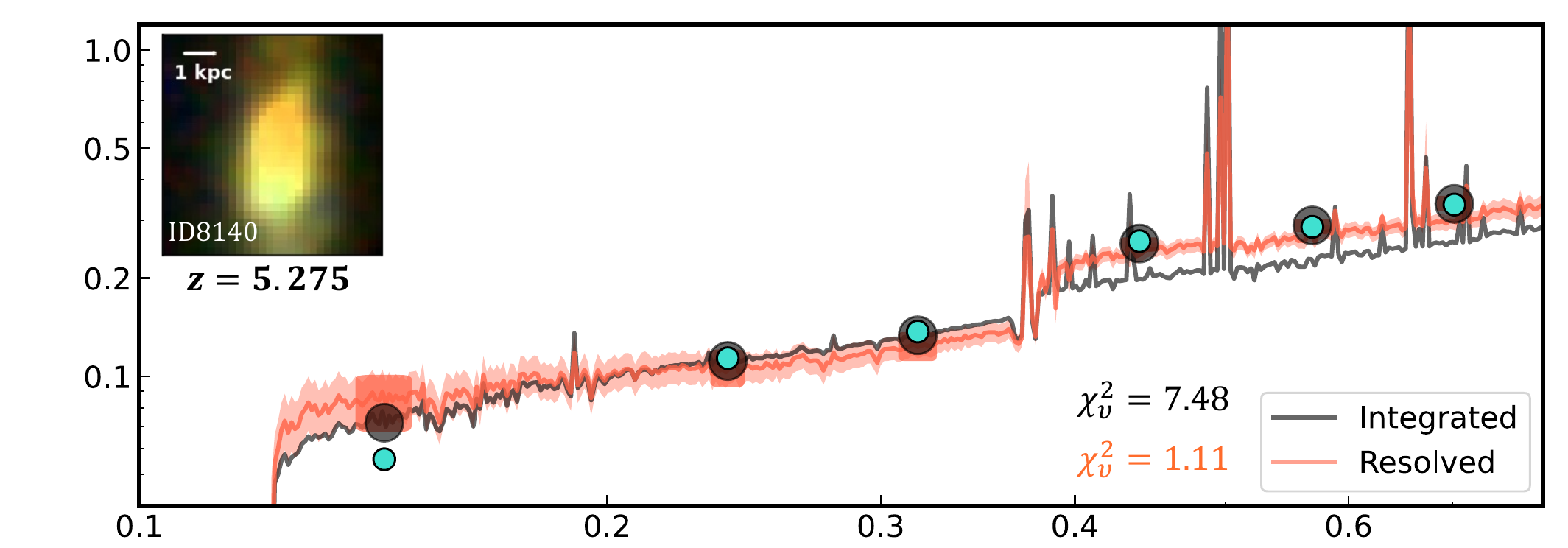}
\includegraphics[width=\columnwidth]{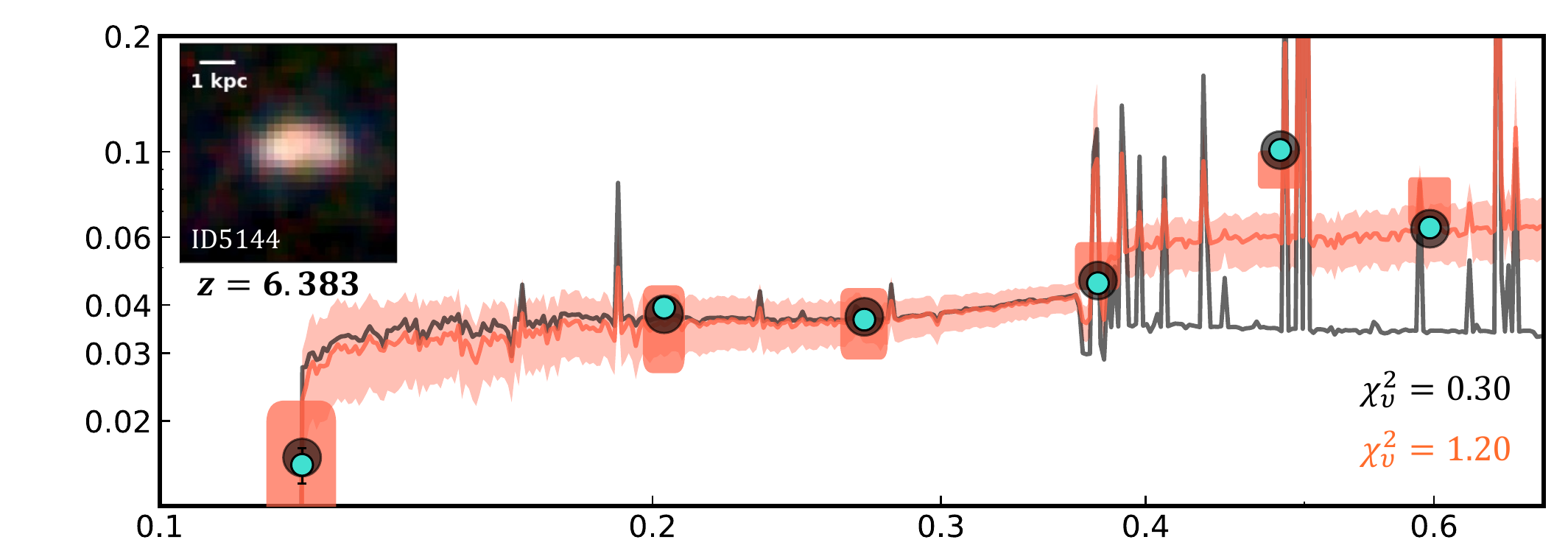}
\includegraphics[width=\columnwidth]{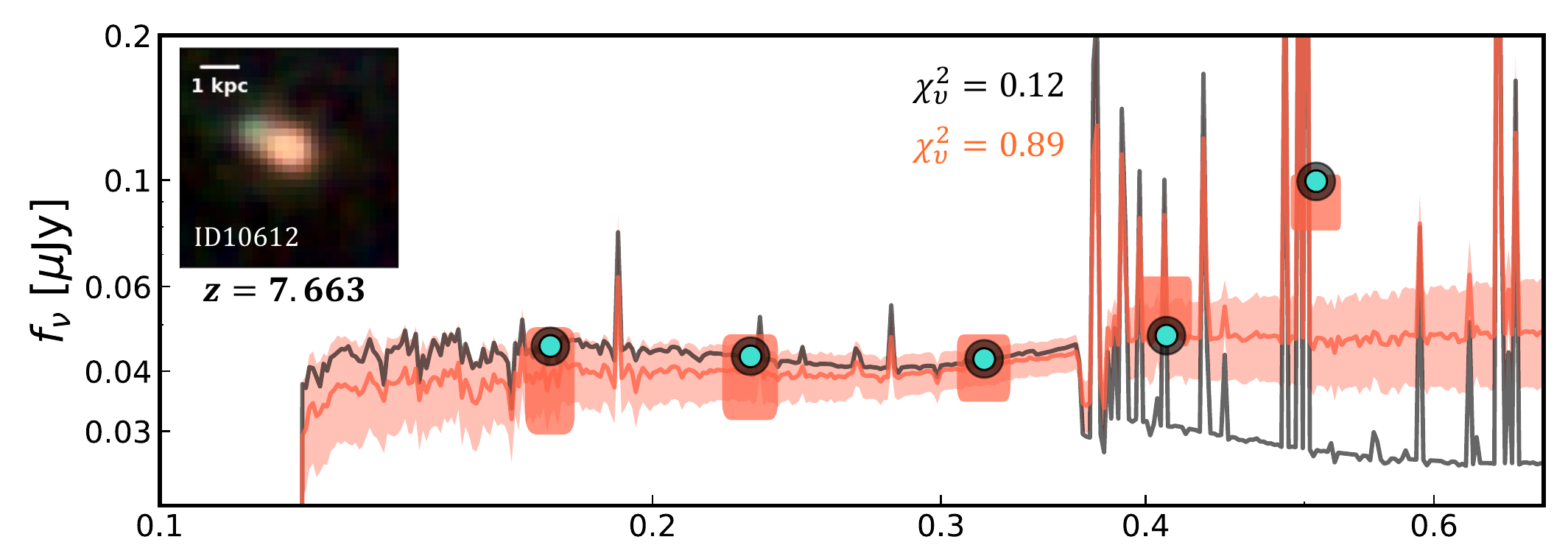}
\includegraphics[width=\columnwidth]{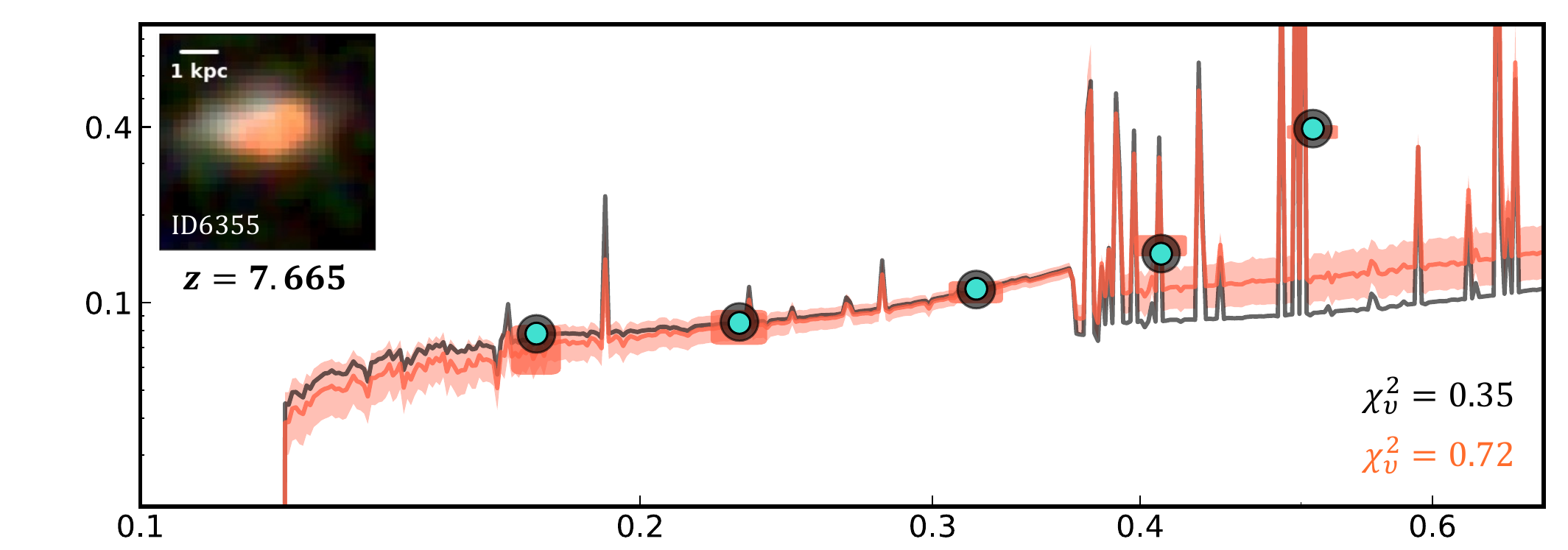}
\includegraphics[width=\columnwidth]{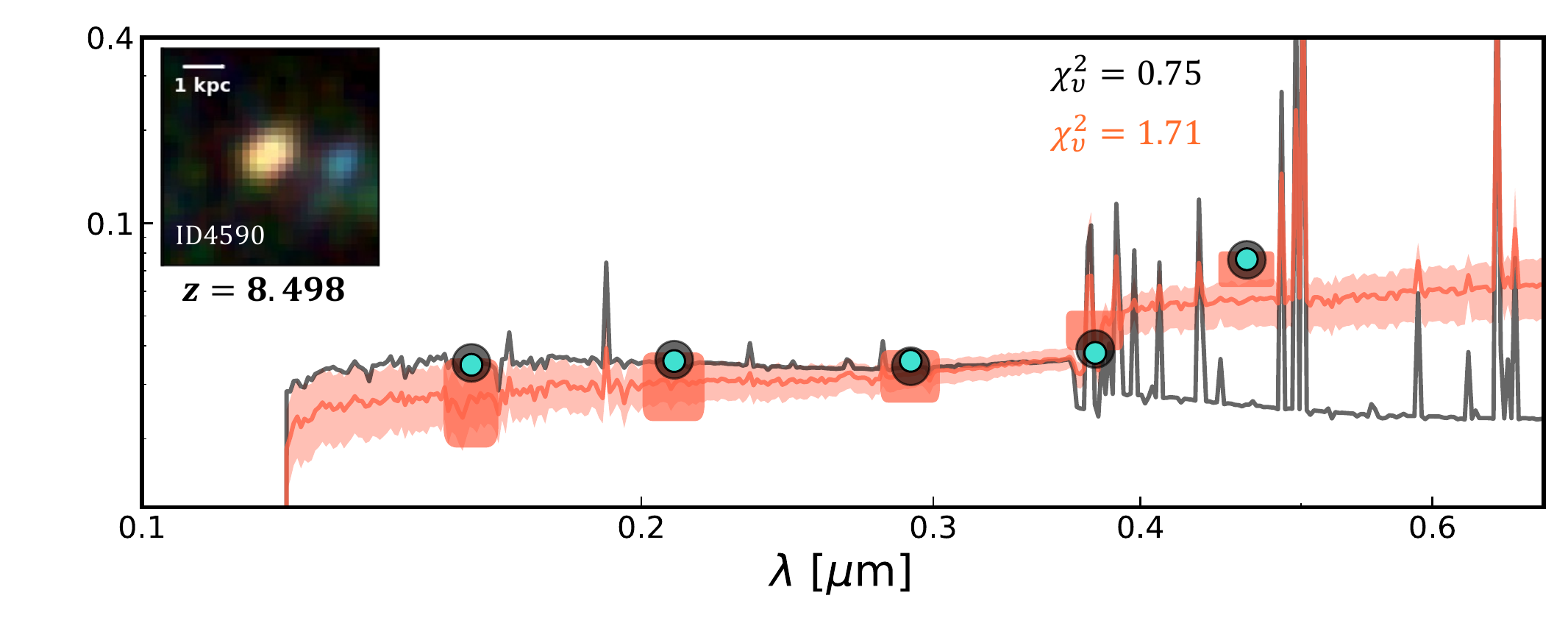}
\caption{Best fit SEDs and models for the integrated (black curve and circles) and resolved (red curve and squares) modelling of the five galaxies studied in this work. The turquoise points and errorbars correspond to the integrated NIRCam photometry. The red curve is inferred by summing the posterior distributions in all pixels, and calculating the 50th percentile of the resulting one. The shaded regions correspond to the 16th and 84th percentile of the summed posterior distribution. The inset cutouts correspond to the same RGB images as in Figure~\ref{fig:cutouts}. The reduced $\chi^2$ values of each fit are indicated.} 
\label{fig:integrated_seds}
\end{figure}

To perform this test, we sum the photometry in each observed band for the pixels that fulfil our S/N criteria, so that we only consider the same pixels that we fit in the spatially resolved analysis shown in Figures~\ref{results_z5.275}--\ref{results_z8.498}. With the sum of the photometry in each filter, we then use \textsc{Bagpipes} to find the best fit SED and infer the integrated physical parameters. We use the exact same set up as for the spatially resolved run, described in \S\ref{sec:sed_bagpipes}. We present the integrated fits with the best fit SEDs in Figure~\ref{fig:integrated_seds}. The inferred integrated physical properties for each galaxy can be found in Table~\ref{tab:integrated_full}, in Appendix \ref{app:integrated}. In Figure~\ref{fig:integrated_seds}, we can see that both the resolved and integrated models (red and black curves, respectively) fit adequately the photometry (turquoise points), with reduced $\chi^2$ values within the range $0.1-7.5$. The surprising finding is that both best fit SEDs are considerably different. For all galaxies, we find that the high equivalent widths that we could spatially locate in the resolved analysis within a clump, now completely dominate the overall fit. This results in inferring extremely young ages in the integrated light, and potentially too low stellar masses as a result.

\begin{figure}[!t]
\centering
\includegraphics[width=\columnwidth]{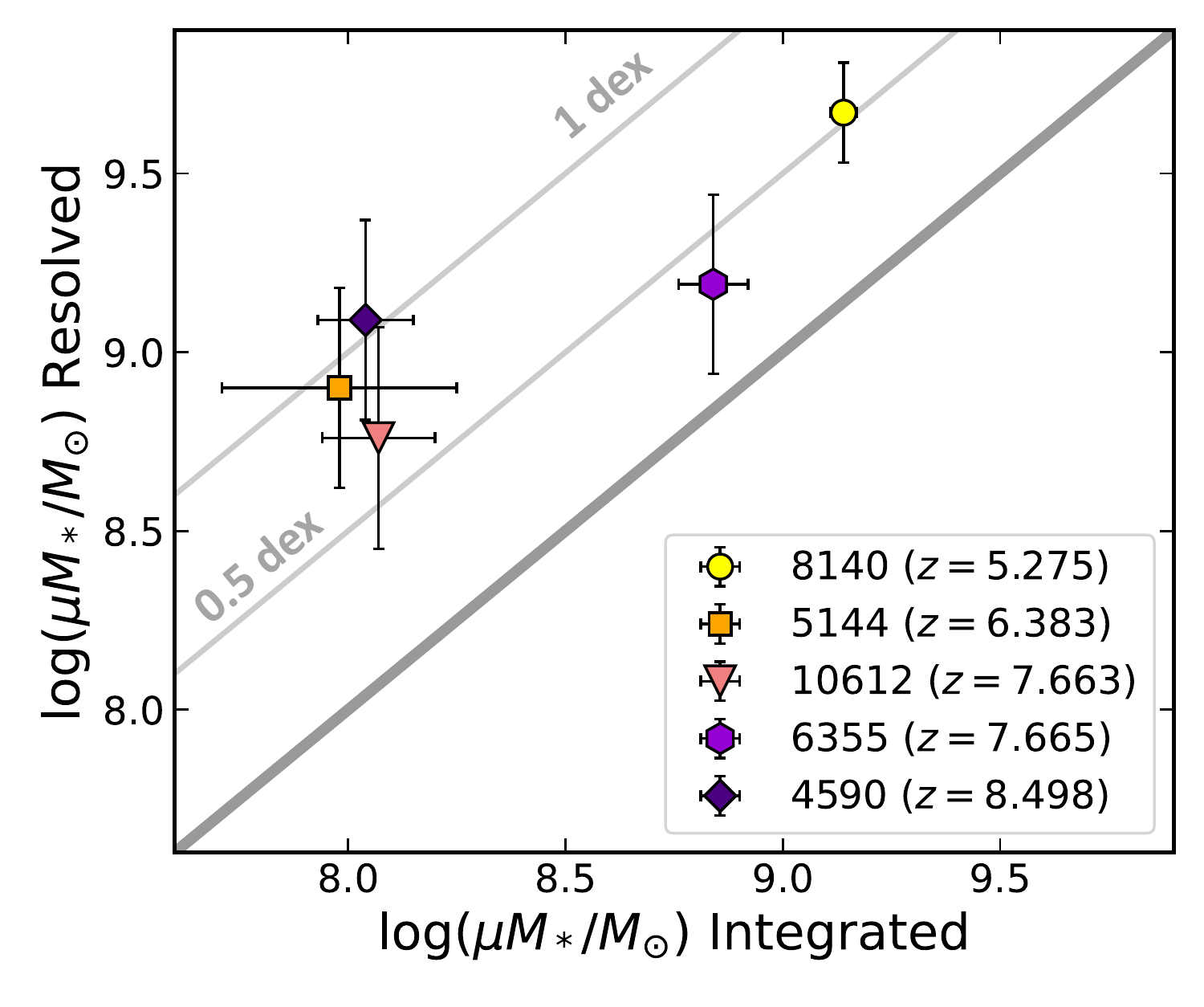}
\caption{Comparison between the stellar mass that we infer in the spatially resolved analysis, versus the integrated fit. No lensing correction is applied in any of the two estimates. The plotted values are shown Table~\ref{tab:integrated}. The one-to-one line is indicated, as well as the 0.5 and 1 dex offset lines.} 
\label{fig:integrated_sr}
\end{figure}

Figure~\ref{fig:integrated_sr} shows the comparison between the stellar mass estimates that we infer in the spatially resolved analysis and the integrated fit. Table~\ref{tab:integrated} provides these values, as well as the mass weighted ages. We obtain the spatially resolved mass estimate by summing the stellar mass inferred in each individual pixel. The resolved average age is inferred as the mean mass weighted age of all pixels. In both cases, the resolved uncertainties reported in Table~\ref{tab:integrated} are calculated with the 16th and 84th percentile of the summed posterior distribution over all pixels. For the integrated run, the output from \textsc{Bagpipes} is directly reported, and the uncertainties are calculated with the 16th and 84th percentiles from the posterior distribution. We find that the integrated run estimates systematically lower stellar masses than the spatially resolved one, from $\sim$0.5 up to 1 dex lower, seemingly without any trend with the redshift. This, as explained above, is a consequence of being forced to choose extremely young stellar populations to fit the integrated light, due to the strong emission lines dominating it completely. We see this in Table~\ref{tab:integrated}, since all galaxies have an integrated age of under 10~Myr, except the one at lowest redshift, with a best fit age of $14.9^{+1.7}_{-1.0}$~Myr. We demonstrate this by fixing the age in \textsc{Bagpipes} to the average stellar age inferred in the spatially resolved analysis. By doing this, the estimate of the stellar mass in the integrated run increases, retrieving closer values to the spatially resolved masses. Moreover, we see that the average mass weighted ages in the resolved analysis are all $>10$~Myr, and significantly older than the integrated ages, even when considering the large uncertainties associated with the age estimate.

\begin{table}[t]
\centering
\begin{tabular}{c c c c c c}
\hline
\multirow{2}{*}{$z$} & \multirow{2}{*}{ID} & \multicolumn{2}{c}{log($\mu M_*/M_{\odot}$)} & \multicolumn{2}{c}{Age [Myr]} \\
 & & Integrated & Resolved & Integrated & Resolved  \\ 
\hline
 5.275 & 8140 & $9.14^{+0.03}_{-0.02}$ & $9.67^{+0.10}_{-0.14}$ &$14.9^{+1.7}_{-1.0}$ & $251^{+172}_{-139}$  \\
 6.383 & 5144 & $7.98^{+0.27}_{-0.11}$ & $8.90^{+0.20}_{-0.28}$&$2.5^{+1.1}_{-1.3}$ & $147^{+138}_{-103}$ \\
 7.663 & 10612 & $8.07^{+0.13}_{-0.08}$ & $8.76^{+0.28}_{-0.31}$&$1.6^{+0.7}_{-0.5}$& $71^{+85}_{-51}$  \\
 7.665 & 6355 & $8.84^{+0.08}_{-0.06}$ & $9.19^{+0.25}_{-0.22}$&$1.3^{+3.6}_{-0.2}$ & $25^{+38}_{-18}$   \\
 8.498 & 4590 & $8.04^{+0.11}_{-0.06}$ & $9.09^{+0.23}_{-0.28}$&$1.9^{+0.4}_{-0.5}$ & $111^{+108}_{-76}$  \\
\hline
\end{tabular}
\caption{Values for the stellar mass and mass weighted stellar age that we infer with \textsc{Bagpipes}, both in the integrated run and the spatially resolved analysis presented in this work. We plot the mass values in Figure~\ref{fig:integrated_sr}.}
\label{tab:integrated}
\end{table}

\begin{figure}[!t]
\centering
\includegraphics[width=0.95\columnwidth]{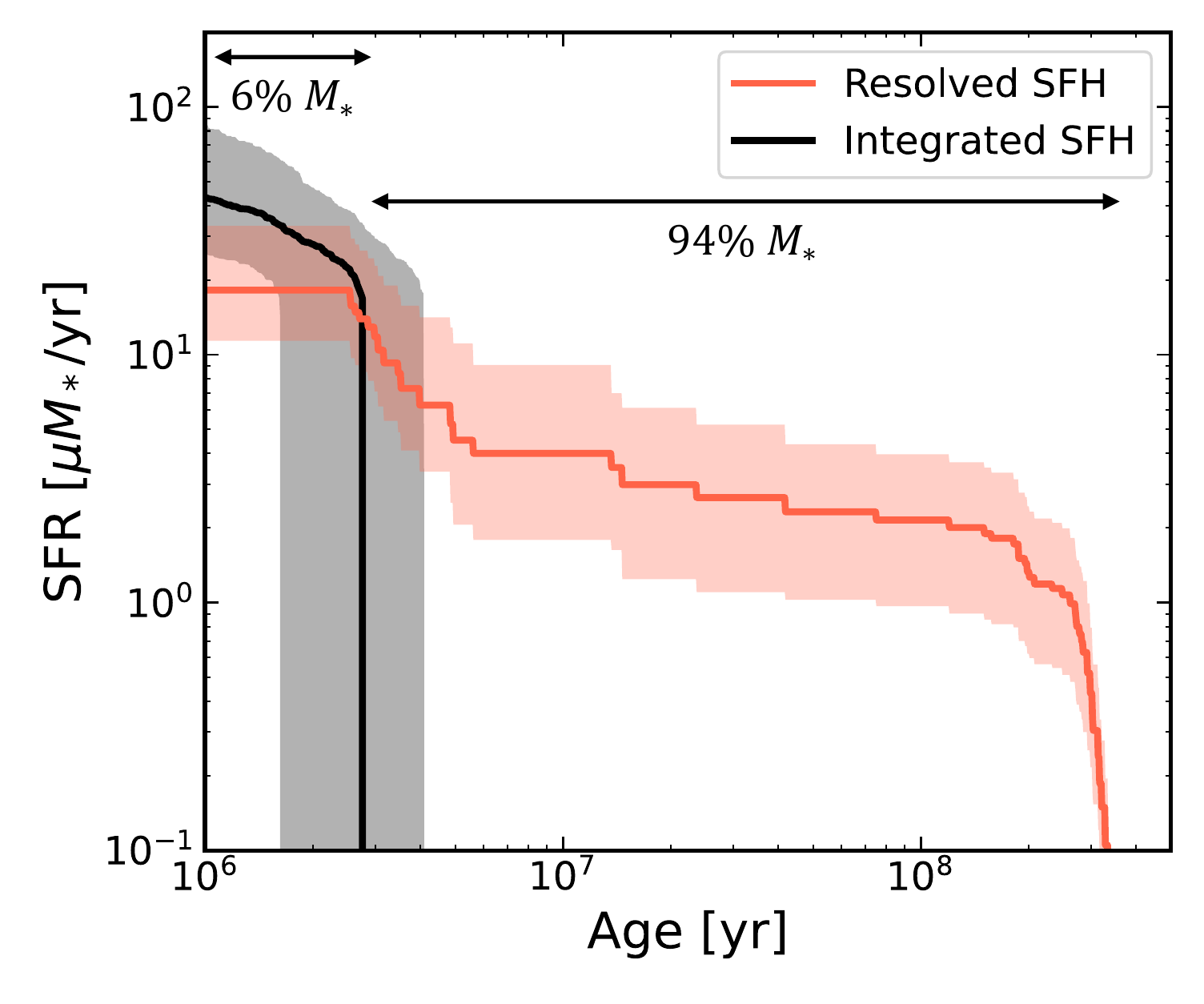}
\caption{Comparison between the star formation history that we infer in the spatially resolved analysis (red curve), versus the integrated fit (black curve), for the galaxy ID10612 at $z=$~7.663. The shaded regions correspond to the 16-84th percentile range in each case.} 
\label{fig:integrated_sr_sfh}
\end{figure}

Figure~\ref{fig:integrated_sr_sfh} shows how the SFH affects the inferred stellar mass. We plot the sum of the SFH inferred for the spatial pixels, as well as the SFH estimated in the unresolved analysis, for the galaxy ID10612 at $z=7.663$. The integrated SFH consists of a single burst with very young age ($\sim2$~Myr), whereas the spatially resolved SFH is a distribution that covers a wider age range, reaching up to $\sim300$~Myr. For this galaxy, this would mean a formation redshift of $z\sim12$. We see that, whereas the integrated analysis forms all stellar mass within less than 10~Myr, this only corresponds to $\sim6\%$ of the spatially resolved stellar mass, directly proving where the mass discrepancy is coming from. We obtain the same results in the SFH comparison for all galaxies studied in this work.

This could considerably change our current picture of mass assembly in the early Universe, particularly while our samples are limited to the brightest candidates in small numbers. These systematics would affect from the stellar mass functions that we have derived so far at high redshifts, to our cosmological models of galaxy formation and mass build-up, since all our observations and mass estimates at high redshift until now have been based on integrated measurements. Overall, the nature of these galaxies can completely change by having a resolved picture.

On top of this, we adopt a parametric constant SFH model. Some works have shown that using instead a non-parametric SFH model can lead to inferring larger stellar masses by up to 1 dex (all in an integrated approach), in particular for galaxies like the ones studied here, where a burst with very young stellar ages dominates the integrated light \citep[e.g.,][]{Leja19,Lower20,Tacchella22,Whitler22}. Non-parametric SFHs therefore seem like the better option when considering integrated SEDs, since we have shown that there can be significant spatial variation in star formation (see Figure~\ref{fig:integrated_sr_sfh}). Moreover, works like \cite{Bisigello19} argue that we need to include the two MIRI bands in order to constrain better the stellar mass estimates in high redshift galaxies, particularly in young galaxies with nebular emission lines, such as the ones we study here.

\subsection{Comparison with Other Works}

To put our results into context, we compare the physical parameters that we infer with other published works on these targets. As stated before, only integrated analyses are available in the literature. We expect our integrated measurements and estimates of the physical properties to be lower than any other work, since aperture photometry yields larger fluxes on all bands ($\sim20\%$ larger for the galaxy ID10612 at $z=7.663$ if we use a 0\farcs5 aperture), given that we only consider pixels that fulfil our S/N criteria. On top of this, previous works have used varying data reduction, zero points, SED fitting codes, SFH models, attenuation curves, magnification corrections, amongst other differences. Therefore, here we mostly focus on comparing the physical nature of the sources, as inferred by different works.

The mass-weighted ages that we infer in our integrated analysis are all consistent within the uncertainties with those found for these same five targets by \cite{Carnall22}, from which we reproduce the SED fitting assumptions and parameters, except using a differently reduced photometry, a \cite{Calzetti00} attenuation curve, as well as extending our nebular grid as explained in \S\ref{sec:method}. 

\cite{Tacchella22} studies the stellar populations of three of our targets, the ones at highest redshifts $z=7.663$, $z=7.665$ and $z=8.498$ (IDs 10612, 6355 and 4590, respectively). They use \textsc{Prospector} with a flexible non-parametric SFH prescription instead to model the SEDs. They find that the highest redshift galaxy (ID4590) is undergoing a recent burst, inferring a young stellar age under 10~Myr, like we obtain in our integrated analysis. They cannot rule out older stellar ages in their analysis. In our spatially resolved maps (see the age map in Fig.~\ref{results_z8.498}), we find that most pixels have a mass weighted age above $\sim100$~Myr, except the centrally located young burst, which dominates the integrated light. The particularly striking case is the $z=7.665$ galaxy (ID6355). \cite{Tacchella22} also infer an extremely young age of $3^{+29}_{-1}$~Myr, and they rule out the presence of older stellar populations, since the extreme [\ion{O}{3}]+H$\beta$ lines dominate the emission. In this work we find a shell of older stars, confirmed by the empirical colour gradients as well as the other tests discussed in \S\ref{section:caveats} and \S\ref{app_7665}. Finally, for the target ID10612 at $z=7.663$, they argue that its SFH together with its morphology, could indicate that this source is undergoing a merger. This is consistent with what we find, both in terms of the empirical colour gradients, as well as the clumpiness of the blue F150W and F200W bands. Moreover, we find two distinct populations within the galaxy in the spatially resolved maps (see Figure~\ref{results_z7.663}). Consistent with our results here, they find that inferring older stellar ages (in their case by adding emission line constraints with NIRSpec spectra) leads to larger stellar masses of up to 1 dex for the galaxy ID4590 at $z=8.495$, which is exactly what we find in our integrated versus spatially resolved comparison. 

Our work confirms the issue discussed in \cite{Tacchella22,Topping22} and \cite{Whitler22}, where young stars outshine and dominate the emission when compared to older stars, the presence of which is difficult to rule out via integrated measurements. This leads to inferring lower stellar masses. In \cite{Tacchella22} they conclude that the SFH prior is of vital importance, and they only infer stellar ages older than 10~Myr in their fits when using a non-parametric SFH model with a continuous prior and older populations present. This leads to an increase in the stellar masses of up to 0.6~dex. \cite{Whitler22} also use various SFH models to explore the potential presence of old stellar populations in seemingly young galaxies. They find stellar masses larger by up to an order of magnitude with non-parametric SFH versus constant SFH models. The impact of the SFH model in the inferred stellar mass has been studied by other works \citep[see e.g.,][]{Leja19, Suess22}, with similar results.

Outshining and its effects on stellar mass estimates have been studied at lower redshifts \citep[see e.g.,][]{Maraston10,Pforr12}. Our results agree with previous works such as \cite{Sorba18}, where they find a discrepancy in the inferred stellar masses of up to a factor of $\sim5$, when having resolved SED fitting, albeit their study only reaches $z=2.5$. Moreover, they propose that unresolved studies should apply corrections to their mass estimates. This resolves the mass missing problem, in which a tension is found between the observed stellar mass density of the Universe and the star formation rate density \citep[see e.g.,][]{Leja15}. By correcting stellar mass functions with resolved estimates, they find that these agree better with the observed star formation densities collected by \cite{Madau14}. \cite{Leja20} also solve this discrepancy by using flexible non-parametric SFH priors, which produce older ages, thus inferring $\sim50$\% higher stellar mass density.

\cite{Endsley22} study a population of UV-faint galaxies at a similar redshift range $z\sim6.5-8$, also finding that the SEDs are dominated by young stellar populations, exhibiting low masses. They find the majority of their objects to appear very blue ($\beta\sim-2$), with some dusty galaxies ($\beta\sim-1$). With our integrated analysis, we find three galaxies with $\beta\sim-2$, and two targets with a value closer to $-1$. In the spatially resolved maps, we find values $-2<\beta<-1$, with some very dusty regions reaching values around $\beta\sim-0.5$. 

At this redshift range, the majority of targets with high EW([\ion{O}{3}]+H$\beta$) have an inferred young stellar age when considering a constant SFH, just as we find in our integrated analysis \citep[see Fig. 9 in][]{Endsley22}. On the other hand, there are works that find evolved stellar populations ($>$100~Myr) at even higher redshifts, such as \cite{Furtak22} with $z\sim10-16$ candidates in the SMACS0723 field, and \cite{Leethochawalit22} with $7<z<9$ photometrically-selected galaxies in the GLASS-JWST ERS program, which infer a median mass-weighted age of 140~Myr.

From a resolved point of view, in a sample of $z\sim 6-8$ galaxies in the Extended Groth Strip (EGS) field, \cite{Chen22} find multiple clumps dominated by young stellar populations, as well as significant variations in the equivalent width of the [\ion{O}{3}]+H$\beta$ lines. They find EWs with extreme values such as the ones we find for most of our targets (of the order $\sim300-3000$~Å), which also yield young ages in their fits, confirming once more what we are finding for these high redshift targets. Moreover, \cite{PerezGonzalez22} also find strong [\ion{O}{3}]+H$\beta$ emission in a spatially resolved analysis using \textit{HST} and \textit{JWST} data from the CEERS survey in the EGS. They also link these findings with very young starburst with possibly an underlying older stellar population.

In summary, our results are consistent with the works that have been published so far studying these same galaxies, or targets at a similar redshift range with \textit{JWST}. By integrating our maps, we can produce similar results and draw equivalent conclusions to the integrated works performed so far in these sources. By producing a spatially resolved analysis, we can demonstrate the presence of underlying older stellar populations that are otherwise outshined in the integrated analyses, inferring larger stellar masses and considerably affecting our picture of the nature of these high redshift galaxies.

\subsection{Caveats} \label{section:caveats}

As briefly mentioned before, one could argue that the ``shells" of older stellar populations where the young stars with high EWs are embedded, could be instead a result of the dust-age degeneracy present in SED fitting softwares. To test if the gradient in age is real, we perform a test in which we fix the extinction to the value given by \cite{Tacchella22}. We find a similar gradient, where there is a shell of older stellar populations surrounding the bursty young star forming region. The effects of dust and age are now blended into the age map, since we fix the $A_V$, but the gradient persists. On top of that, the individual fits that we obtain fixing the dust obscuration are very poor, compared to leaving $A_V$ as a free parameter in the \textsc{Bagpipes} fit. This gives us confidence that our fits with $A_V$ as a free parameter sample better the galaxy properties, and the gradient observed is real. In Appendix~\ref{app_7665}, we discuss in more details the age uncertainties, focusing on the target ID6355 at $z=7.665$.

Another caveat that our spatially resolved analysis could have is whether the process of PSF-matching affects our inferred maps. One could argue that the mass weighted age map could result as an artifact of the PSF-matching procedure, where flux is re-distributed radially to match the resolution of the F444W band. We only observe this radial distribution on the age and EW maps. We observe horizontal gradients on all the rest of physical properties, as well as the empirical colours. The gradients that we observe extend across spatial scales larger than the FWHM of the F444W PSF. We therefore conclude that this is not an effect of PSF-matching, but a true young stellar population clump centrally located in most galaxies.

Finally, besides the S/N threshold that we impose in the noisiest bands, one could still doubt whether there is enough S/N per pixel to be able to infer robust physical parameters. To test this, we apply a Voronoi tesselation binning method on the targets, in order to achieve bins with a constant minimum S/N across the image and filters. Imposing a minimum S/N of 5 or even up to 10 in all bands, we find the same gradients and trends that we observe in the maps of the various inferred physical parameters in all galaxies. This, combined with the fact that our fits display good reduced $\chi^2$ values, gives us confidence that the S/N in each native pixel is sufficient to provide trustworthy estimates.

%%%%%
\section{Summary and Conclusions} \label{sec:conclusions}

We present the first spatially resolved analysis of spectroscopically confirmed $5<z<9$ galaxies in the SMACS0723 ERO field. We use images in 6 bands obtained with NIRCam onboard \textit{JWST}, spanning the wavelength range $0.8-5 \mu$m. With the SED fitting software \textsc{Bagpipes}, we model the spectral energy distributions on a pixel-by-pixel basis, being able to infer the physical parameters on a $180-240$ parsec scale. Our main findings and conclusions are the following:

\begin{itemize}
    \item All galaxies are resolved and display strong empirical colour gradients. Even at these early times, these galaxies display multiple star forming clumps.
    
    \item We find regions that exhibit high EW of the [\ion{O}{3}]+H$\beta$ emission (up to $\sim3000-4000$~Å). These extreme starbursts are embedded within regions with less specific star formation, which points to very bursty star formation happening on small scales ($<1$~kpc), not galaxy-wide.
    
    \item The strong line emission regions dominate the integrated light, biasing the fits towards very young inferred ages of the stellar population ($<10$~Myr). Only a resolved analysis demonstrates the presence of older stellar populations, which can be seen in the spatial maps.
    
    \item Resolving the stellar populations on a pixel-by-pixel basis leads to inferring from 0.5 up to $\sim1$~dex larger stellar masses, when compared to an integrated analysis. Our analysis extends previous findings on the problem of outshining and its effects on stellar mass estimates, which so far has only been studied at lower redshifts (up to $z\sim3$).

\end{itemize}

Current and upcoming observations with \textit{JWST} will allow us to characterise the early Universe and first galaxies in a new and more complete way. The combination of having confirmed redshifts with NIRSpec, and the unprecedented resolution and depth of NIRCam imaging, will transform how we study galaxies, changing our current views on their internal structure and mass assembly, amongst others. The systematics in stellar mass estimates found in this work would have strong implications in the shape and evolution of the stellar mass function at high redshift, particularly while samples are limited to small numbers of the brightest candidates. Furthermore, the process of galaxy formation could be more extended and earlier than previously thought, as is implied by the presence of evolved older stellar populations being outshone by the youngest stars. Only with a spatially resolved analysis, we can begin to untangle the complexity of the internal structure of galaxies at this epoch.

\begin{acknowledgments}

The Cosmic Dawn Center is funded by the Danish National Research Foundation (DNRF) under grant DNRF140. This work is based on observations made with the NASA/ESA/CSA \textit{James Webb Space Telescope}. The data were obtained from the Mikulski Archive for Space Telescopes at the Space Telescope Science Institute, which is operated by the Association of Universities for Research in Astronomy, Inc., under NASA contract NAS 5-03127 for \textit{JWST}. These observations are associated with program ID 2736, as part of the Early Release Observations. P.O. is supported by the Swiss National Science Foundation through project grant 200020\_207349. This work received funding from the Swiss State Secretariat for Education, Research and Innovation (SERI). C.A.M acknowledges support by the VILLUM FONDEN under grant 37459. S.F. acknowledges the support from NASA through the NASA Hubble Fellowship grant HST-HF2-51505.001-A awarded by the Space Telescope Science Institute, which is operated by the Association of Universities for Research in Astronomy, Incorporated, under NASA contract NAS5-26555. Cloud-based data processing and file storage for this work is provided by the AWS Cloud Credits for Research program. 

\facilities{\textit{JWST} (NIRCam)}
\software{Astropy \citep{2013A&A...558A..33A,2022ApJ...935..167A}, Matplotlib \citep{Hunter:2007}, NumPy \citep{numpy}, SciPy \citep{scipy}, \texttt{grizli} \citep{Brammer19,grizli,grizli22}}
\end{acknowledgments}

\clearpage

\appendix \label{sec:appendix}

\section{Age Uncertainty} \label{app_7665}

\begin{figure}[ht]
\centering
\includegraphics[width=\columnwidth]{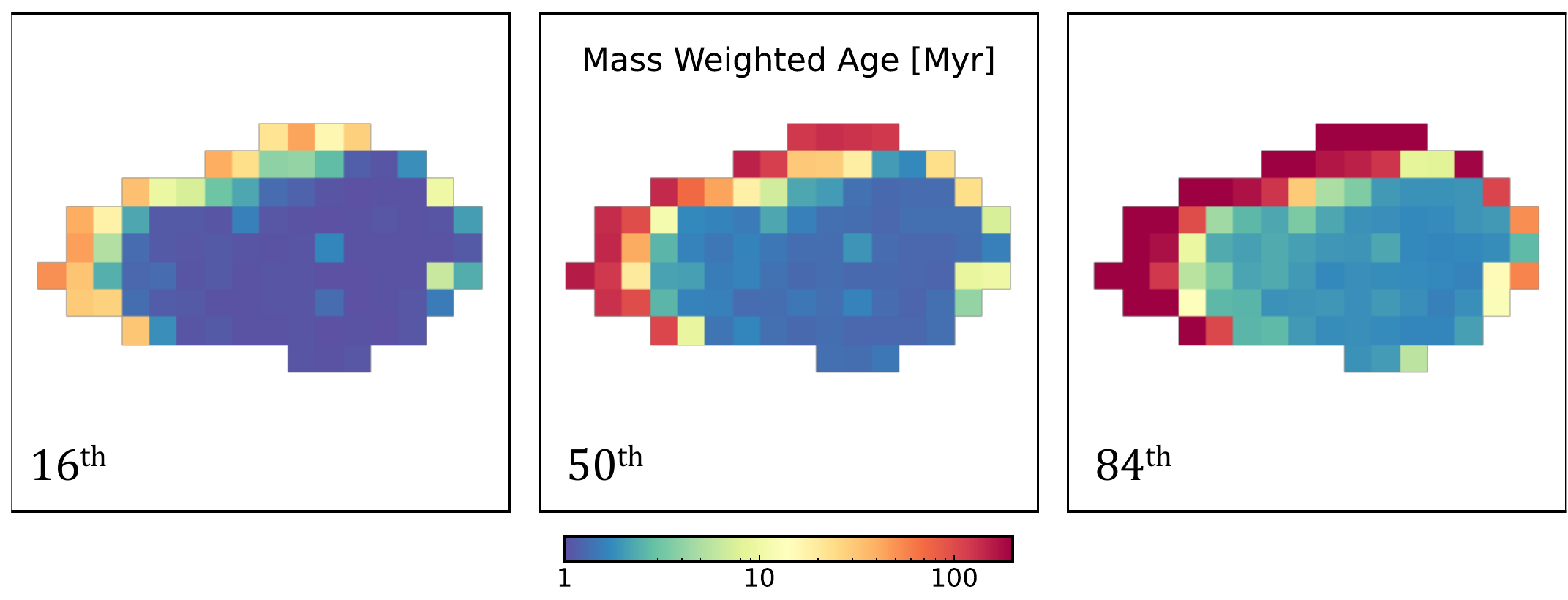}
\caption{Mass weighted age 16th, 50th and 84th percentiles of the posterior distribution inferred with \textsc{Bagpipes} on the galaxy ID6355 at $z=7.665$.} \label{fig:age_unc}
\end{figure}

As discussed in \S\ref{section:caveats}, the degeneracy between age and dust could yield uncertain estimates. This could be particularly concerning for the source ID6355 at $z=7.665$, and one could argue that the shell that we see on the age map is not real. This would mean that there is no underlying older stellar population being outshined by the young stellar population that dominates the emission for this galaxy. Figure~\ref{fig:age_unc} displays the 16th, 50th and 84th percentiles of the mass weighted age, obtained with the posterior distribution that \textsc{Bagpipes} infers for each individual pixel. In the 50th percentile, which is the one we choose to display for every inferred physical property in Figures~\ref{results_z5.275}--\ref{results_z8.498}, we see a shell of older stellar ages, as discussed in \S\ref{sec:results_SR}. With the 16th percentile image, we see that even if we assume the maximum lower uncertainty, the shell of old stars would still display ages above 10~Myr. That is still older than what is inferred by other works on this target, as well as in our integrated analysis, where we infer an age of $1.3^{+3.6}_{-0.2}$~Myr. With the 84th percentile image, we see that these old stars could be up to hundreds of Myr old. Therefore, even within the uncertainty range, we can confidently say that there are older stellar components present in this galaxy, opposite to what is concluded by \cite{Tacchella22} and \cite{Carnall22}. This is only visible with a careful spatially resolved analysis. On the other hand, the very extended region where the EW is extremely high, can only be fit by young stellar templates. Considering the uncertainties, we still only obtain young stellar populations in that region.

\section{Integrated Properties} \label{app:integrated}

Table~\ref{tab:integrated_full} shows the resulting physical parameters inferred with \textsc{Bagpipes} on the integrated fit for each galaxy. As a reminder, we expect these values to be lower than the ones inferred by other works, since we only consider the pixels here that fulfil a certain S/N criteria, instead of performing aperture photometry, which would yield larger fluxes and different physical estimates. The integrated run is performed like this to be able to do a one-to-one comparison with the spatially resolved analysis, as discussed in \S\ref{sec:integrated}.

\begin{table*}
\centering
\begin{tabular}{l c c c c c}
\hline
\multirow{2}{*}{Integrated Properties} & ID8140 & ID5144 & ID10612 & ID6355 & ID4590 \\ 
 & $z=5.275$ &$z=6.383$ & $z=7.663$ & $z=7.665$ & $z=8.498$ \\
\hline
log($\mu M_*/M_{\odot}$) & $9.14^{+0.03}_{-0.02}$ & $7.98^{+0.27}_{-0.11}$ & $8.07^{+0.13}_{-0.08}$ & $8.84^{+0.08}_{-0.06}$ & $8.04^{+0.11}_{-0.06}$ \\
SFR [$\mu M_{\odot}$/yr] & $16.0^{+1.3}_{-0.7}$ & $1.0^{+0.5}_{-0.2}$ & $1.2^{+0.3}_{-0.2}$ & $6.9^{+1.3}_{-0.8}$ & $1.1^{+0.3}_{-0.2}$ \\
$A_V$ & $1.16^{+0.02}_{-0.02}$ & $0.58^{+0.18}_{-0.12}$ & $0.45^{+0.11}_{-0.09}$ & $1.01^{+0.07}_{-0.06}$ & $0.48^{+0.12}_{-0.08}$ \\
Age [Myr] & $14.9^{+1.7}_{-1.0}$ & $2.5^{+1.1}_{-1.3}$ & $1.6^{+0.7}_{-0.5}$ & $1.3^{+3.6}_{-0.2}$ & $1.9^{+0.4}_{-0.5}$ \\ 
EW([\ion{O}{3}]+H$\beta$) [Å] & $729\pm44$ & $1996\pm386$ & $3048\pm329$ & $3884\pm252$ & $2565\pm254$ \\
UV slope $\beta$ & $-1.20\pm0.03$ & $-1.88\pm0.11$ & $-2.08\pm0.09$ & $-1.40\pm0.05 $ & $-2.03\pm0.09$ \\
\hline
\end{tabular}
\caption{Resulting physical properties inferred with \textsc{Bagpipes} for the integrated measurements of the five galaxies.}
\label{tab:integrated_full}
\end{table*} 

\bibliographystyle{aasjournal}

\end{document}